\documentclass{article}

\usepackage{PRIMEarxiv}
\usepackage{amsmath}
\usepackage[utf8]{inputenc} 
\usepackage[T1]{fontenc}    
\usepackage{hyperref}       
\usepackage{url}            
\usepackage{booktabs}       
\usepackage{amsfonts}       
\usepackage{nicefrac}       
\usepackage{microtype}      
\usepackage{lipsum}
\usepackage{fancyhdr}       
\usepackage{graphicx}       
\graphicspath{{media/}}     

\title{EVOLVE: Predicting User Evolution and Network Dynamics in Social Media Using Fine-Tuned GPT-like Model
}

\author{
  Ismail Hossain, Md Jahangir Alam, Sai Puppala, Sajedul Talukder \\
  School of Computing \\
  Southern Illinois University Carbondale, IL, USA, 62901\\
  \texttt{\{ismail.hossain, mdjahangir.alam, saimaniteja.puppala, sajedul.talukder\}@siu.edu} \\
}

\begin{document}
\maketitle

\maketitle
\begin{abstract}
Social media platforms are extensively used for sharing personal emotions, daily activities, and various life events, keeping people updated with the latest happenings. From the moment a user creates an account, they continually expand their network of friends or followers, freely interacting with others by posting, commenting, and sharing content. Over time, user behavior evolves based on demographic attributes and the networks they establish. In this research, we propose a predictive method to understand how a user evolves on social media throughout their life and to forecast the next stage of their evolution. We fine-tune a GPT-like decoder-only model (we named it E-GPT: Evolution-GPT) to predict the future stages of a user's evolution in online social media. We evaluate the performance of these models and demonstrate how user attributes influence changes within their network by predicting future connections and shifts in user activities on social media, which also addresses other social media challenges such as recommendation systems.

\end{abstract}

\section{Introduction}
\label{sec: introduction}

In the dynamic landscape of online social media, the evolution of user behavior stands as a pivotal area of study. With the advent of advanced algorithms, user-generated content, and intricate network structures, understanding the trajectory of users within these digital ecosystems has become more crucial than ever. Our research delves into this captivating realm, exploring the next phase of evolution for users in online social media platforms.


Over the years, researchers have meticulously dissected various facets of online social media, shedding light on user engagement patterns, network dynamics, and the impact of algorithms on information dissemination~\cite{bakshy2015exposure,pariser2011filter,granovetter1973strength}. Evolution in social networks like community evolution tracking~\cite{tracking_dakiche_2019}, evolution of individual social power in a social network~\cite{evolution_ye_2018}, evolution of social networks in response to life cycle events~\cite{predicting_sharmeen_2015}, evolution because of creating social link as an effect of triadic closure~\cite{role_weng_2013}, and temporal evolution of an online location-based social network~\cite{evolution_allamanis_2012}. Some facts are responsible for the evolution of users on social evolution like cultural diversity~\cite{evolution_grabowski_2006} as well as changes in individuals' roles and social status in the network~\cite{analyzing_lin_2009}.

\begin{figure}
    \centering
    \includegraphics[width=0.45\textwidth]{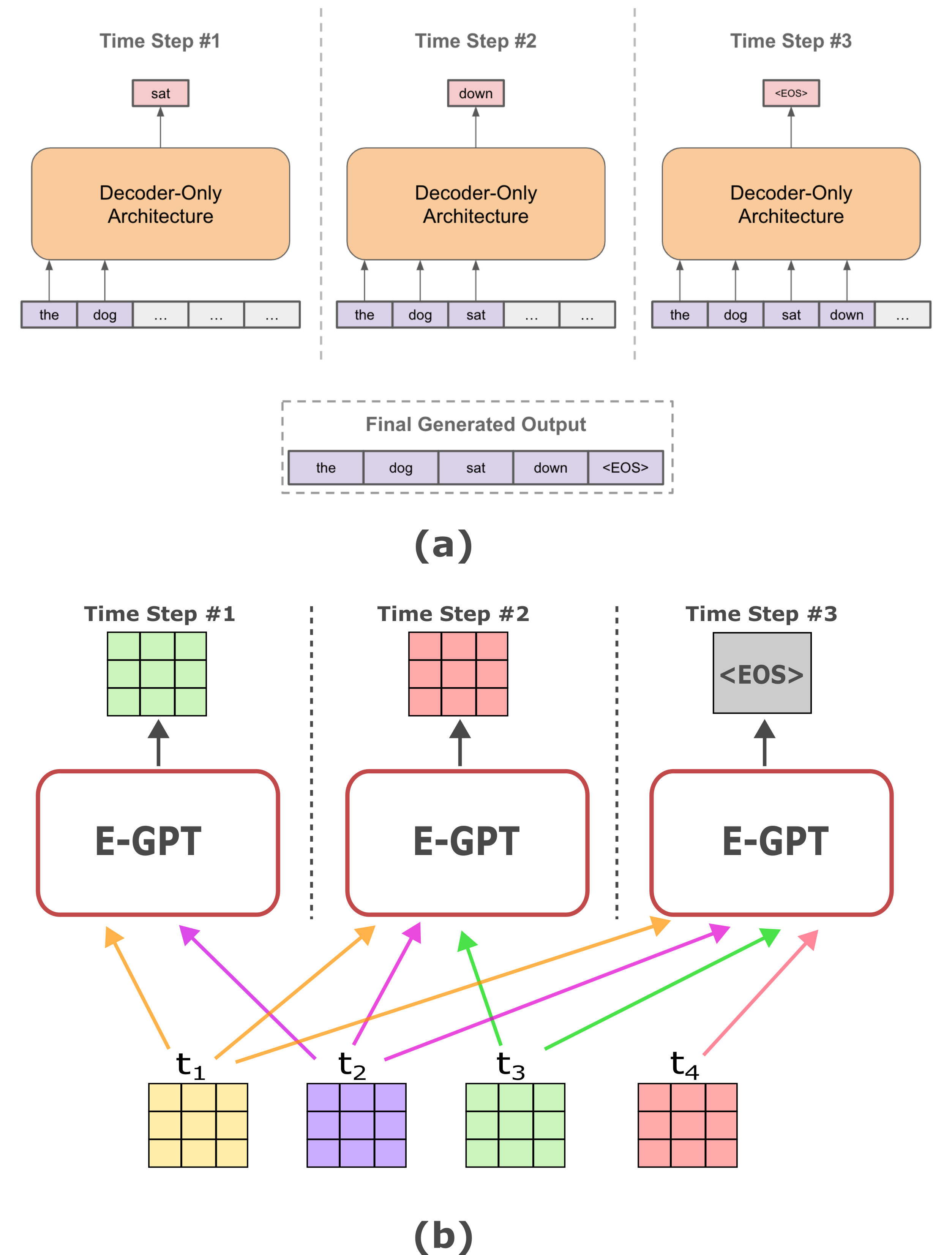}
    \caption{(a) Autoregressive output from a decoder-only transformer architecture~\cite{wolfe2022} (b) Temporal data a user in social media predicting the next stage of the evolution.}
    \label{fig:decoder-only-gpt}
\end{figure}

\begin{figure*}[h]
    \centering
    \includegraphics[width=\textwidth]{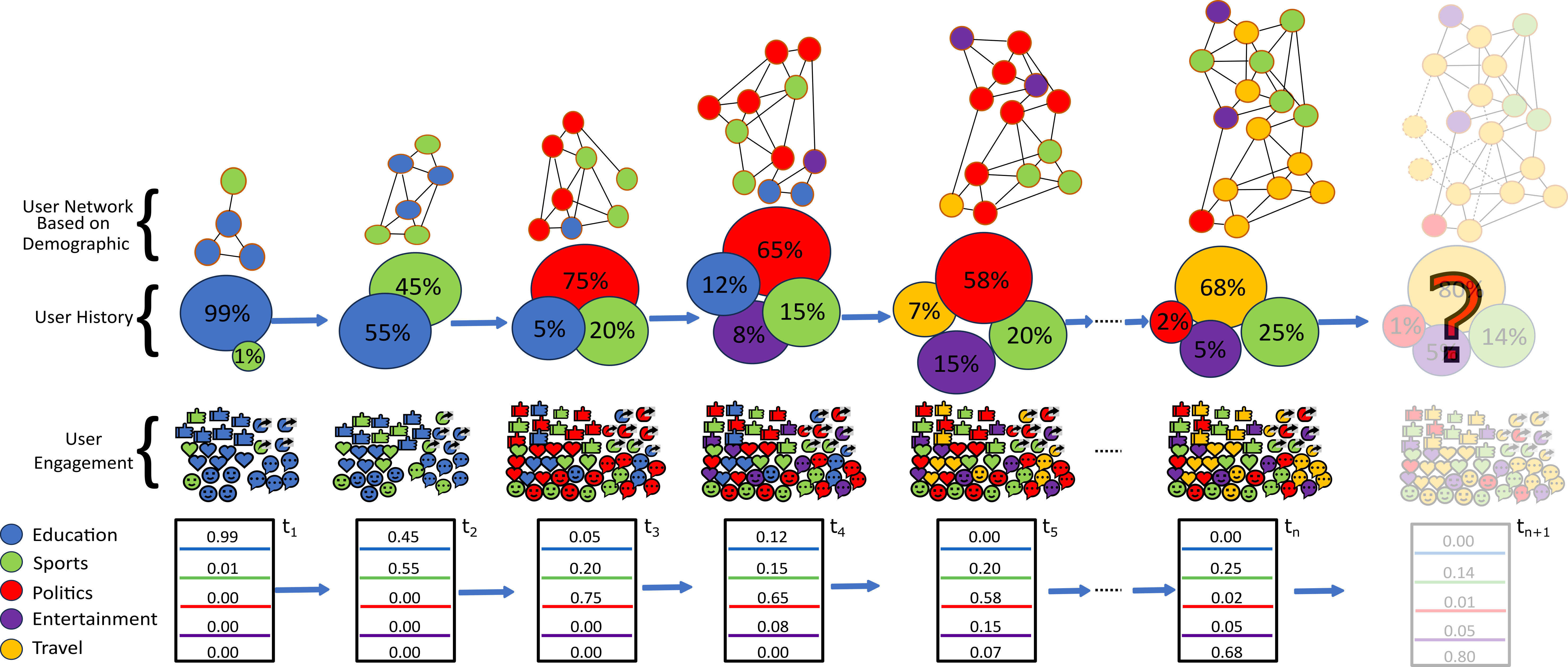}
    \caption{System architecture of user evolution in online social media}
    \label{fig: system-architecture}
\end{figure*}

Our approach adopts a system by incorporating the user network, demographic data, history, and engagement to comprehensively analyze the next phase of user evolution in online social networks. By synthesizing diverse methodologies and leveraging cutting-edge techniques, we aim to uncover nuanced patterns and emergent phenomena that underpin the evolution of user behavior.


At the heart of our research lies a fundamental question: How do users evolve within the intricate web of social media platforms like Facebook, Twitter, Reddit, and so on, and what factors drive this evolutionary process? We seek to unravel the underlying mechanisms governing user adaptation, exploring how individual behaviors, network structures, and platform dynamics interplay to shape the trajectory of users over time. By elucidating the drivers of user evolution, we aim to provide insights that can inform the design of more adaptive and user-centric social media platforms.

Text generation using GPT models typically involves predicting the next word or sequence in a given context. This inspired us to explore the evolution of user behavior on social media platforms. Figure~\ref{fig: system-architecture} illustrates how a user's social media presence evolves based on various factors. As users engage in social media, their network changes and their posting habits shift. For instance, initially, a user might predominantly post about Education (99\%) and rarely about Sports (1

Rather than predicting text, our goal is to forecast the future state of a user’s social media presence, including their network and activities such as posting or interacting with others. From the inception of a social media account, a user's social status evolves, influenced by factors like changing interests, educational qualifications, occupations, and social networks. This dynamic evolution can be modeled similarly to text sequences.

To achieve this, we applied GPT architectures to predict the next stage in a user’s social media evolution. We aim to forecast changes in their network and activity patterns using the sequential prediction capabilities of these advanced language models.

GPT (decoder-only transformer) is a versatile model known for its skill in predicting the next item in a sequence~\cite{radford2019language}. This ability enables it to perform tasks such as text translation, question answering, and summarization~\cite{radford2019language}. Specifically, the GPT-like model utilizes the decoder portion of the Transformer architecture.

\textbf{Why the Decoder?} The choice of the decoder in language models is intentional. The decoder's masked self-attention layers ensure the model only considers preceding tokens when generating a representation for the current token, preventing it from seeing future tokens. This is crucial for language modeling as it maintains an autoregressive approach, where the output at time \(t\) becomes the input at time \(t+1\). This setup enables the model to continuously predict the next token in a sequence, as shown in Figure~\ref{fig:decoder-only-gpt}(a).


In this paper, we embark on a journey to dissect the next phase of user evolution in social media platforms as we show in the Figure~\ref{fig:decoder-only-gpt}(b). We begin by providing a comprehensive review of existing literature, laying the groundwork for our empirical analysis. Subsequently, we describe our motivation in Section~\ref{sec: motivation}, state the problem in Section~\ref{sec: problem-formulation}, and outline our research methodology in Section~\ref{sec: methedology}, detailing the data sources, analytical techniques, and theoretical frameworks employed in our study. We then present our findings, elucidating key insights gleaned from our analysis, and show our experimental results in Section~\ref{sec: experiment}. We mention our limitation in Section~\ref{sec: discussion}. Finally, in Section~\ref{sec: conclusion}, we conclude by discussing the implications of our research and outlining avenues for future exploration in this burgeoning field. Through this structured approach, we aim to contribute to a deeper understanding of user evolution in social media and pave the way for more informed decision-making in the realm of digital communication.

In summary, we introduce the following contributions:
\begin{itemize}
    \item \textbf{Modeling Social Evolution Using Demographics, History, and Engagement:}
    We propose a model that predicts online social evolution for a user based on their activities and interactions within social media.
    
    \item \textbf{Developing a Sequential Evolution Framework:}
    We create a sequential framework to analyze users' historical data and predict future changes, thereby understanding the drivers of social evolution.
    
    
    \item \textbf{Utilizing GPT Architecture for User Evolution Prediction:}
    We adopt the GPT model to generate predictions about the future state of a user in online social media, considering user networks and various attributes.
\end{itemize}

\subsection{Research Objectives}
\label{sec:model:rq}

In this article, we investigate the following research questions on user evolution in social media:

\begin{itemize}
\item
\textbf{(RQ1)}: Can we propose a system that will predict user evolution in social media?
\item
\textbf{(RQ2)}: Can the system measure the evolution in terms of user data that drives this evolution?

\item
\textbf{(RQ3)}: How effective is the GPT architecture at accurately predicting user evolution in social media, especially when the data is ambiguous or multifaceted?

\item
\textbf{(RQ4)}: What challenges and opportunities arise when implementing dynamic changes to predict evolution?

\item
\textbf{(RQ5)}: What are the real-world applications of predicting user evolution in social media?
\end{itemize}

Our ultimate goal with this research is to develop a system capable of identifying user evolution in social media, which can be leveraged for important applications such as recommendation systems.

\section{Literature Review}
User evolution in social media refers to how users' behaviors, preferences, and interactions change over time. Prior studies have examined various aspects of this evolution, including engagement patterns, content creation, and network dynamics.

\textbf{Initial Adoption Phase}
When users first join Twitter, they typically explore the platform, follow a few accounts, and start posting tweets. Their interactions are limited as they are still getting accustomed to the platform's features. This phase has been characterized as a period of low activity and exploration \cite{Zhao2016}.
New users often consume content passively, reading tweets from the accounts they follow without engaging much \cite{Yang2011}.
Profiles may be sparse, with minimal information and a few tweets.

\textbf{Engagement Growth Phase}
As users become more familiar with Twitter, they start following more accounts and engage more actively by liking, retweeting, and replying to tweets. They begin to participate in trending topics and use hashtags. This increased engagement is documented as users move from lurkers to more active participants \cite{Joinson2008}.
Users start creating more content, sharing their thoughts, opinions, and experiences. They may also begin to curate their profile with more detailed bios and profile pictures \cite{Lampe2006}.
Users’ follower count starts to grow as they interact more, leading to increased visibility \cite{Burke2009}.

\textbf{Personalization and Customization Phase}
Users develop specific interests and tailor their feed by following accounts that align with their preferences. They may also mute or unfollow accounts that are no longer relevant to them \cite{Kwak2010}.
Engagement becomes more selective and meaningful. Users participate in discussions that interest them and contribute more substantial content, such as threads or media-rich posts \cite{Li2014}.
Users optimize their profiles to reflect their personal brand or identity, including pinning important tweets and using cover photos \cite{McCay-Peet2017}.

\textbf{Influence and Authority Phase}
Some users evolve into influencers or thought leaders within their niche. They gain a significant following and their tweets receive high engagement. Research has shown the importance of this phase in shaping online communities and trends \cite{Marwick2011}.
Content becomes more strategic, with a focus on maintaining and growing their audience. Influencers may collaborate with brands or other users and participate in high-profile discussions \cite{Cha2010}.
These users have a broad network and can influence trends and conversations on the platform \cite{Jin2014}.

\textbf{Mature User Phase}
Mature users may reduce the frequency of their posts but maintain a consistent presence. They focus on quality over quantity and engage in more meaningful interactions \cite{Bakshy2011}.
They consume content from a curated list of trusted sources and engage with content that aligns with their established interests and values \cite{Brandtzaeg2009}.
Mature users often contribute to community building by supporting new users, participating in group discussions, and sometimes moderating communities \cite{Tang2012}.

\textbf{Implications for Recommendation Systems}
As users evolve, recommendation systems need to adapt to their changing preferences and behaviors. Early recommendations might focus on onboarding content, while later recommendations can become more refined and personalized \cite{Kumar2010}.
Understanding user evolution helps in predicting future engagement patterns, allowing the system to suggest content that aligns with the user’s current interests and potential future trends \cite{Ricci2015}.
Analyzing how users' networks expand and change over time can help in recommending new connections, groups, or communities that may be of interest to the user \cite{jannach2010recommender}.

By studying user evolution, social media platforms can enhance user experience, improve content relevance, and foster stronger community engagement.

\section{Problem Formulation}
\label{sec: problem-formulation}
To address \textbf{RQ2}, we designed our system based on user data that drive user evolution in social media. Below, we detail the formulation process of our proposed system.
Let $U$ be the set of users and $A$ be the adjacency matrix that represents the connection between two users if both are online social media friends or following either one and demographic data can be defined by $D$, where $D = \{a,g,o,l\}$, here, {a: age, g: gender, o: occupation, l: location, and $H$ as user history based on post categories $C = \{c_1, c_2, c_3, ..., c_k\}$ and $E$ for user engagement which consists of user interaction with other users' post, including reaction, sharing, commenting. Feature vector $X$ consists of demographic information, user history, and user engagement. Suppose, we have total $T$ times $X$ for a single user.

Let's, create a combined feature matrix for each time step 
$t \in \{1,2, \dots, T\}$. One approach is to concatenate the adjacency matrix and demographic matrix, history matrix, and engagement matrix for $N$ users to have feature embedding $F$.

$F^{(t)} = \left[ A^{(t)} \| D^{(t)} \| H^{(t)} \| E^{(t)} \right]$

\[
A^{(t)}_{ij} = 
\begin{cases}
    1 & \text{if connection exists between users $u_i$ and 
    $u_j$} \\
    0 & \text{otherwise}
\end{cases}
\]

\textbf{Input:} Construct a sequence of combined feature matrices over the $T$ time steps: $S = \left[ F^{(1)}, F^{(2)}, F^{(3)}, \dots, F^{(T)} \right]$
\textbf{Output:} A Evolution model that estimates the probability
that the users with evolution history $\left[ F^{(1)}, F^{(2)}, F^{(3)}, \dots, F^{(T)} \right]$ will be $F^{(T+1)}$ at the $(T + 1)$-th step.

\section{Motivation}
\label{sec: motivation}
GPT is currently among the most widely used language models for text generation and prediction. Numerous experiments have demonstrated their exceptional performance in generating coherent and contextually relevant text sequences. Text generation typically involves predicting the next word, sentence, or sequence in a given context. This inspired us to explore the evolution of user behavior on social media platforms.

To address \textbf{RQ1}, rather than focusing on text prediction, our aim is to predict the future state of a user’s social media presence, including their network and probable activities such as posting content or interacting with others. From the inception of a social media account, a user's social status evolves over time. For instance, someone who starts with 10 friends might have 200 friends after a few years. Additionally, user activities on social media change due to various factors such as evolving thoughts, educational qualifications, occupations, and social networks. This dynamic change is a sequence that can be modeled similarly to text sequences.

Recognizing this, we applied the architectures of GPT to predict the next state in a user’s social media evolution. By doing so, we aim to forecast changes in their network and activity patterns, leveraging the sequential prediction capabilities of these advanced language models.

\begin{figure*}
    \centering
    \includegraphics[width=0.8\textwidth]{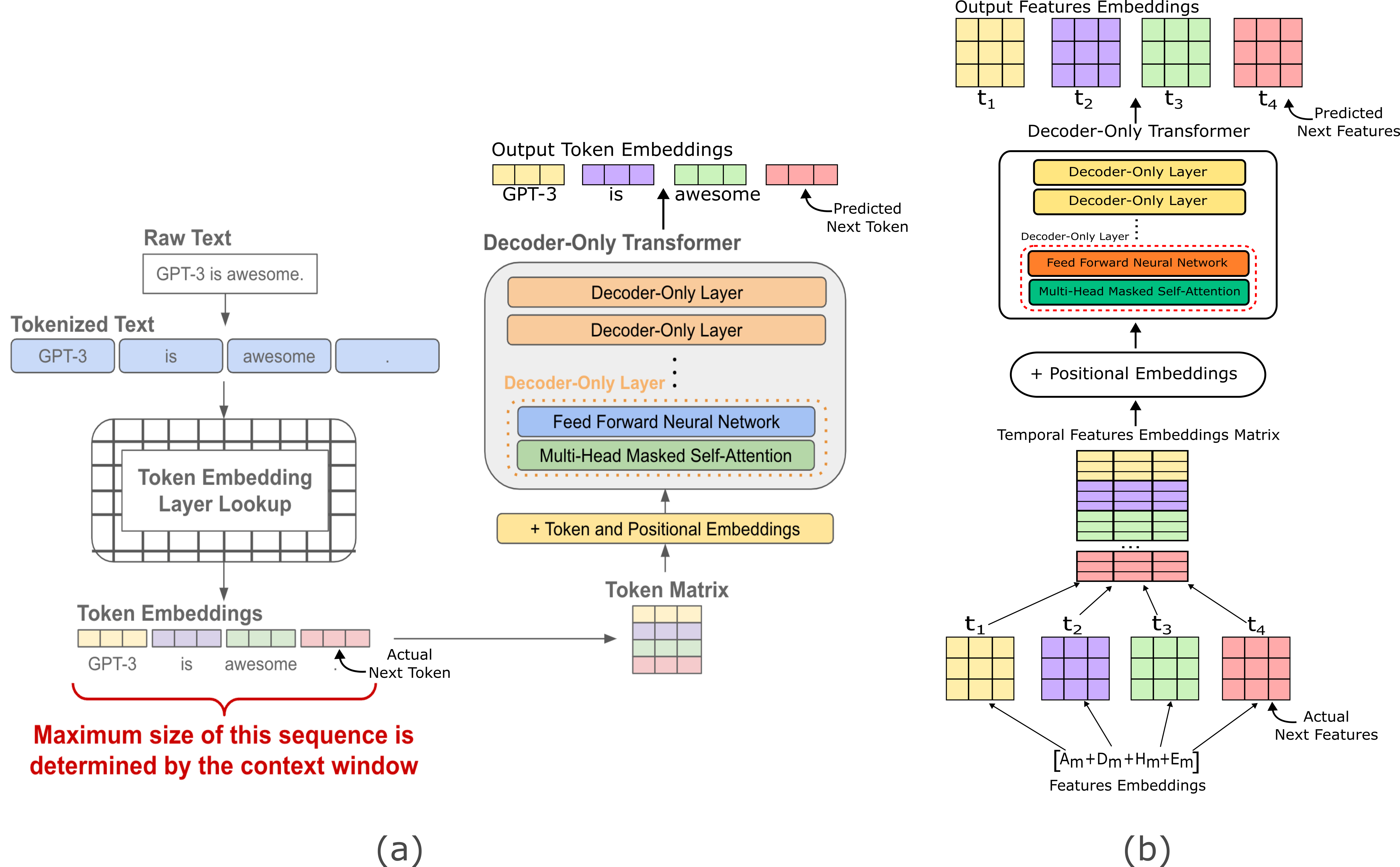}
    \caption{(a) Decoder language model architecture~\cite{wolfe2023}, (b) Our Decoder architecture}
    \label{fig:decoder-only-gpt-archi}
\end{figure*}

\section{Methodology}
\label{sec: methedology}
Figure~\ref{fig:decoder-only-gpt-archi} illustrates the temporal feature embeddings, which are analogous to the sequence of words in a sentence and that address the \textbf{RQ3}. In this figure, $t_1, t_2$, and $t_3$ represent previous stages of evolution, while $t_4$ represents the next evolution stage, marked as both the actual and predicted features. Unlike token embeddings used for next-word prediction, we use 2D feature embeddings as input. Our E-GPT model predicts 2D output feature embeddings. From these predictions, we can extract the user's network, demographic attributes, history, and engagement. The predicted results provide us with a user adjacency matrix that reflects the attributes of all users. Below, we detail the model architecture and how we designed the model for this study.

\subsection{Adapting E-GPT Architecture for 2D Sequences}

Given the combined feature matrix $\mathbf{F}^{(t)}$ at time step $t$, where $\mathbf{F}^{(t)} \in \mathbb{R}^{N \times (N + d + h + e)}$ ($N$ is the number of users and $d$ is the number of demographic features, $h$ is the number of history feature, and $e$ is the number of engagement feature), we adapt the E-GPT architecture as follows:

\subsection{Input Embedding}
Map the combined feature matrix to a higher-dimensional space using an embedding function $\mathbf{E}$:
\begin{equation}
\mathbf{H}_0^{(t)} = \mathbf{E}(\mathbf{F}^{(t)}) \in \mathbb{R}^{N \times d_h}
\end{equation}
where $d_h$ is the hidden dimension.

\subsection{Positional Encoding}
Add a learnable positional encoding $\mathbf{P}^{(t)} \in \mathbb{R}^{N \times d_h}$ to retain temporal and spatial order:
\begin{equation}
\mathbf{H}_0^{(t)} = \mathbf{H}_0^{(t)} + \mathbf{P}^{(t)}
\end{equation}

\subsection*{3. Flattening for Transformer Input}
Concatenate the sequence of embedded matrices $\{\mathbf{H}_0^{(1)}, \mathbf{H}_0^{(2)}, \ldots, \mathbf{H}_0^{(T)}\}$ along the temporal dimension and reshape it into a format suitable for the Transformer:
\begin{equation}
\mathbf{H}_{\text{input}} = \text{reshape}(\{\mathbf{H}_0^{(1)}, \mathbf{H}_0^{(2)}, \ldots, \mathbf{H}_0^{(T)}\}) \in \mathbb{R}^{(T \cdot N) \times d_h}
\end{equation}




\subsection{Transformer Decoder for GPT}
The reshaped input $\mathbf{H}_{\text{input}}$ is processed through $L$ layers of Transformer decoders. Each layer consists of masked multi-head self-attention followed by a position-wise feed-forward network:

For $l = 1$ to $L$:
\begin{align}
\mathbf{H}_{l}^{\prime} &= \text{MaskedMultiHead}(\mathbf{H}_{l-1}) + \mathbf{H}_{l-1} \\
\mathbf{H}_{l} &= \text{FFN}(\mathbf{H}_{l}^{\prime}) + \mathbf{H}_{l}^{\prime}
\end{align}

where $\text{MaskedMultiHead}$ is the masked multi-head attention mechanism, and $\text{FFN}$ is the feed-forward network.

The final encoded representation is:
\begin{equation}
\mathbf{H}_{\text{output}} = \mathbf{H}_{L} \in \mathbb{R}^{(T \cdot N) \times d_h}
\end{equation}

\subsection{Reshaping Back to 2D}
Reshape the output back to the 2D structure over time:
\begin{equation}
\mathbf{H}_{\text{reshaped}} = \text{reshape}(\mathbf{H}_{\text{output}}) \in \mathbb{R}^{T \times N \times d_h}
\end{equation}

\subsection{Output Layer}
Pass the reshaped output through a final linear layer to map it back to the original feature space:
\begin{equation}
\mathbf{F}_{\text{pred}}^{(t)} = \mathbf{W}_{\text{out}} \mathbf{H}_{\text{reshaped}}^{(t)} + \mathbf{b}_{\text{out}} \in \mathbb{R}^{N \times (N + d + h + e)}
\end{equation}
where $\mathbf{W}_{\text{out}} \in \mathbb{R}^{d_h \times (N + d + h + e)}$ and $\mathbf{b}_{\text{out}} \in \mathbb{R}^{(N + d + h + e)}$ are learnable parameters.

If we want to get predicted results in terms of adjacency matrix, demographic, history, and engagement, we apply the following equations:
\begin{equation}
\mathbf{A}_{\text{pred}}^{(t)} = \mathbf{F}_{\text{pred}}^{(t)}[:, :N]
\end{equation}
where $\mathbf{F}_{\text{pred}}^{(t)}[:, :N]$ represents the first $N$ columns of $\mathbf{F}_{\text{pred}}^{(t)}$.

Similarly, we have the following:
\begin{equation}
\mathbf{D}_{\text{pred}}^{(t)} = \mathbf{F}_{\text{pred}}^{(t)}[:, N:N+d]
\end{equation}

\begin{equation}
\mathbf{H}_{\text{pred}}^{(t)} = \mathbf{F}_{\text{pred}}^{(t)}[:, N+d:N+d+h]
\end{equation}

\begin{equation}
\mathbf{E}_{\text{pred}}^{(t)} = \mathbf{F}_{\text{pred}}^{(t)}[:, N+d+h:N+d+h+e]
\end{equation}

\section{Experiment}
\label{sec: experiment}

\subsection{Dataset}
\label{subsec:dataset}
Sometimes, domain-specific datasets are not readily available or open to the public, making it difficult to collect the data needed for research purposes~\cite{guo2024generative}. This addresses \textbf{RQ4}. In our case, constructing users' networks and collecting demographic data, user history, and engagement data is challenging due to data extraction restrictions and associated costs on many social media platforms. We have not found any public dataset that provides all the necessary user information. To continue our study, we experimented with a synthetic dataset. For this study, we need three different types of data: user demographics, user post history, and user post engagement/interaction.

We consider user demographic attributes like age, gender, occupation, and location. User history and engagement are categorized into eight different areas: entertainment, sports, finance, art, education, travel, health, and politics. These are predominant categories on social media, with related posts circulating frequently. Each social media post can fall into one of these eight categories. Additionally, we include user interactions with other posts, such as reactions, sharing, and commenting.

Our synthetic dataset contains a total of 36 features, including three categorical features (age, gender, location), and the rest are numerical. We consider users' age range from 15 to 60, gender as \{M, F, O\}, and occupation as \{physician, teacher, businessperson, actor, engineer, student, sportsperson, writer, banker\}. Locations are chosen from some U.S. cities, such as Dallas and Chicago. Furthermore, in our synthesized dataset, we have users' history percentage for each category and the same for user interactions. For example, a male doctor living in Dallas, mostly connected with other physicians in the city, may post on social media like Facebook about health (71\%), entertainment (5\%), sports (10\%), finance (2\%), art (1\%), education (6\%), travel (3\%), and politics (2\%). These ratios might have been different earlier in life when he was not a doctor or not connected with similar people. Throughout his life, the probabilities of different categories based on his activities might not be the same.

We assigned a total of 2000 posts to each user, with a random number (0 to 10) of posts for each of the 500 users. Additionally, to create an adjacency matrix (size 500 x 500), we randomly assigned binary values of either 0 or 1. Here, 0 means two users do not follow each other or have no connection on social media, and 1 means either one follows the other or both have a connection or are in the same network. Our dataset contains temporal data, meaning each user's data includes 10 different temporal points regarding their network, demographic information, post history, and interaction. Each user's network expands from 10 to 300 users, indicating that the user starts with 10 followers or friends, and this number gradually increases to 300 followers or friends. This scenario reflects real-life user network expansion over time on social media.

\subsection{Model Fine-tune}
We fine-tuned our E-GPT model by experimenting with various hyperparameters: batch sizes (32, 64), learning rates (1e-03, 2e-03, 3e-02), hidden dimensions (64), the number of decoder layers (4-8), and dropout rates (0.1-0.3). Table~\ref{tab:model-performance} presents the model's optimal performance during testing with a batch size of 32, a learning rate of 2e-03, a hidden dimension of 64, 4 decoder layers, and a dropout rate of 0.1. Under these conditions, the model achieved a precision of 69\%, a recall of 75\%, and an F1-score of 72\%. All experiments were conducted over 100 iterations.

\begin{table}
\centering
\caption{E-GPT model performance}
\label{tab:model-performance}
    \begin{tabular}{c|cccc}
        \hline
         Hyper-parameters & Precision & Recall & F1 \\
         \hline
         bs=32,lr=1e-03,hd=64,dl=4,drp=0.1 & 0.5984 & 0.6441 & 0.6204\\
         \hline
         bs=32,lr=2e-03,hd=64,dl=4,drp=0.1 & \textbf{0.6944} & \textbf{0.7511} & \textbf{0.7216}\\
         \hline
         bs=32,lr=3e-02,hd=64,dl=8,drp=0.3 & 0.5542 & 0.5878 & 0.5705\\
         \hline
         bs=64,lr=1e-03,hd=64,dl=4,drp=0.1 & 0.5827 & 0.6518 & 0.6153\\
         \hline
         bs=64,lr=2e-03,hd=64,dl=4,drp=0.3 & 0.4998 & 0.5341 & 0.5163\\
         \hline
         bs=64,lr=3e-02,hd=64,dl=8,drp=0.3 & 0.5614 & 0.5239 & 0.5420\\
         \hline
    \end{tabular}
\end{table}

\begin{table}
\centering
\caption{Network Density and Structural Balance of users' connection for each time step}
\label{tab:density_triadic}
    \begin{tabular}{c|cccccccccc}
        \hline
         ~ & $t_1$ & $t_2$ & $t_3$ & $t_4$ & $t_5$ & $t_6$ & $t_7$  & $t_8$ & $t_9$ & $t_{10}$ \\
         \hline
         Density & 0.5984 & 0.1541 & -0.2757 & 0.6150 & 0.4669 & -0.4242 & 0.1753 & 0.5555 & 0.1753 & 0.5555 \\
         \hline
         Triadic Closure & 0.5984 & 0.1541 & -0.2757 & 0.6150 & 0.4669 & -0.4242 & 0.1753 & 0.5555 & 0.1753 & 0.5555 \\
         \hline
    \end{tabular}
\end{table}

\begin{table}
\centering
\caption{Homophily Effect of different post categories}
\label{tab:user_history}
    \begin{tabular}{cccccccc}
        \hline
        \multicolumn{8}{c}{Assortativity Coefficients for User History} \\
        \hline
         Education & Sports & Politics & Finance & Art & Travel & Health  & Entertainment \\
         \hline
         0.5984 & 0.1541 & -0.2757 & 0.6150 & 0.4669 & -0.4242 & 0.1753 & 0.5555 \\
         \hline
    \end{tabular}
\end{table}

\begin{table}
\centering
\caption{Homophily Effect of different post categories based on user reactions}
\label{tab:user_engage}
    \begin{tabular}{cccccccc}
        \hline
        \multicolumn{8}{c}{Assortativity Coefficients for User Reactions} \\
        \hline
         Education & Sports & Politics & Finance & Art & Travel & Health  & Entertainment \\
         \hline
         0.3051 & 0.2029 & 0.0525 & -0.1379 & 0.1016 & 0.7722 & 0.8255 & -0.9297 \\
         \hline
    \end{tabular}
\end{table}

\begin{table}
\centering
\caption{Homophily Effect of different post categories based on user's comments}
\label{tab:user_engage}
    \begin{tabular}{cccccccc}
        \hline
        \multicolumn{8}{c}{Assortativity Coefficients for User's commenting behavior} \\
        \hline
         Education & Sports & Politics & Finance & Art & Travel & Health  & Entertainment \\
         \hline
         0.3051 & 0.2029 & 0.0525 & -0.1379 & 0.1016 & 0.7722 & 0.8255 & -0.9297 \\
         \hline
    \end{tabular}
\end{table}

\begin{table}
\centering
\caption{Homophily Effect of different post categories based on post sharing}
\label{tab:user_engage}
    \begin{tabular}{cccccccc}
        \hline
        \multicolumn{8}{c}{Assortativity Coefficients for User's sharing behavior} \\
        \hline
         Education & Sports & Politics & Finance & Art & Travel & Health  & Entertainment \\
         \hline
         0.3051 & 0.2029 & 0.0525 & -0.1379 & 0.1016 & 0.7722 & 0.8255 & -0.9297 \\
         \hline
    \end{tabular}
\end{table}

\begin{table}
\centering
\caption{Homophily Effect of User Demographic attributes}
\label{tab:user_demo}
    \begin{tabular}{cccccccc}
        \hline
        \multicolumn{8}{c}{Assortativity Coefficients for User Demography} \\
        \hline
         Age & Gender & Occupation & Location \\
         \hline
         0.9689 & 0.2594 & 0.3862 & -0.9422  \\
         \hline
    \end{tabular}
\end{table}

\subsection{Result}
We experimented with our dataset as described in Section~\ref{subsec:dataset}. This dataset comprises 10-time steps of user data, which includes adjacency matrix, demographic information, post-history, and interactions of the users. We measured Network Density, defined as the ratio of actual connections to possible connections within the network. Table~\ref{tab:density_triadic} presents the network density for previous states ($t_{1-9}$) and the predicted state ($t_{10}$). 

Additionally, we assessed Structural Balance using the score of Triadic Closure~\cite{role_weng_2013,evolution_allamanis_2012}, which indicates the tendency of nodes with shared connections to form additional links. This measure evaluates the model's ability to capture structural balance and transitivity within the network.

Table~\ref{tab:user_history} displays the Assortativity Coefficient calculated with respect to User History data. This coefficient reflects the homophily effect among different post categories, indicating how user history influences the formation of new connections and the tendency of nodes with similar attributes to connect.

Similarly, Table~\ref{tab:user_engage} shows the assortativity coefficient for user engagement, measuring the tendency of nodes with similar behaviors, such as reacting, sharing, and commenting, to form new connections within the network.

Furthermore, Table~\ref{tab:user_demo} illustrates how users with similar demographic attributes tend to connect. The assortativity coefficient score is used here to evaluate the extent to which the model captures homophily~\cite{predicting_sharmeen_2015}.

\section{Discussion and Limitations}
\label{sec: discussion}
Our proposed idea has been implemented using a synthetic dataset specifically generated to include the feature sets required for model fine-tuning. Since this dataset is synthetic, it differs from real-world data, meaning our results would also vary if a real-world dataset were used. Nonetheless, our primary objective is to demonstrate the concept of social evolution over time and to highlight the significance of user demographics, history, and engagement in this process.

We are predicting the trajectory of evolution in online social media, focusing on attributes such as age, gender, occupation, and location. For example, predicting changes in a user's gender over time poses challenges, especially considering the fluidity of gender and the existence of non-binary identities, which were not included in this study. Additionally, while we have focused on certain demographic attributes, other important factors such as educational qualifications, hobbies, religion, ethnicity, and race were not considered. These are crucial for answering \textbf{RQ4}. These attributes could also play a crucial role in understanding social evolution and should be taken into account in future research.

When we talk about predicting user evolution based on user data with social networks in all aspects, we aim to address another problem: recommendation systems. In addressing \textbf{RQ5}, we see the potential for introducing a recommendation system through the prediction evolution process. For example, our goal is to predict the next stage of user evolution in social media. Our model can suggest new friends within the existing user network or users outside of it, functioning similarly to friend recommendations on social media.

Additionally, by considering user demographic attributes, user history, and user engagement, our next-level prediction will recommend posts to users that they have not yet observed.

\section{Ethical Considerations}
We have developed our protocols to collect and analyze our data in an ethical, IRB-approved manner. For analysis purposes, we only stored anonymized data that we collected from the different corpora whose handling does not fall within the PII definition of NIST SP 800-122~\cite{NIST}. Under GDPR~\cite{GDPR}, the use of the information without context, e.g., name or personal identification number, is not considered to be ``personal information''.

\section{Conclusion}
\label{sec: conclusion}
In our study, we illustrate how online social evolution occurs and identify the key attributes driving this evolution. We utilized synthetic datasets to fine-tune the E-GPT model, drawing on the similarities between our concept and next-word or sentence prediction. By examining user data at different time steps—including user adjacency matrices, demographic attributes, history, and engagement—we demonstrated how these evolving datasets drive social evolution. This sequential nature made GPT-like models suitable for predicting the next stage of social media evolution for a user.

We presented results based on three different metrics: network density, structural balance, and assortativity coefficient. These results highlight the significant role of user attributes in forming new connections or reinforcing existing ones within the network. They also show how changes in user activities contribute to social media evolution over time. Additionally, our system addresses another problem in social media, such as recommendation systems.

In the future, we aim to apply our methodology to real-world data to validate our findings. Additionally, we plan to introduce other performance metrics to further evaluate the robustness of our model in capturing the evolution of online social media users.


\end{document}